# IS THERE A DEFICIT OF SOLAR NEUTRINOS?

O. MANUEL
*Nuclear Chemistry, University of Missouri, Rolla, MO 65401 USA*
om@umr.edu

ADITYA KATRAGADA
*Computer Engineering, University of Missouri, Rolla, MO 65401*
adityaa@umr.edu

Abstract. Measurements on the isotopic and elemental compositions of meteorites, planets, lunar samples, the solar wind, and solar flares since 1960 suggest that the standard solar model may be in error. A new solar model suggests that the observed number of solar neutrinos represents at least 87% of the number generated: There is little if any deficit of solar neutrinos.

In 1936 Francis William Aston, who developed the mass spectrograph for accurate determination of atomic weights and expressed his results in terms of nuclear packing fraction, visited Japan to view a solar eclipse. He presented a special lecture that sparked an unusually talented, 19-year old student's interest in nuclear and solar studies, Kazuo Kuroda[1]. Kuroda joined the faculty at the University of Tokyo eight years later, moved to the US after the end of WWII, acquired "Paul" as a first name[1], and correctly predicted natural, self-sustaining, uranium fission reactors in the early history of the Earth[2] and the presence of $^{244}$Pu at the birth of the solar system[3].

In 1956 John Reynolds reported the development of a new, high-sensitivity mass spectrometer for noble gases[4], a significant advancement over earlier instruments. Then in 1960, Reynolds reported two astonishing discoveries with his new mass spectrometer:

- Meteorites contain radiogenic $^{129}$Xe from the *in situ* decay[5] of extinct $^{129}$I.
- The abundance pattern of the other eight stable isotopes of primordial xenon in meteorites is unlike that of terrestrial xenon[6].

The first author (om), a new graduate student in 1960, was called to the office of Professor Kuroda, shown Reynolds' startling results[5,6], and persuaded to use isotopic analyses of noble gases in meteorites as the subject of his PhD research. Manuel went to Reynolds' lab to learn this instrument first hand, but before an instrument could be set up for his graduate work, Fowler et al.[7] concluded that the extinct $^{129}$I in meteorites[5] might have been produced here in the early solar nebula, together with D, Li, Be and B.

Over 40 years of isotope measurements on meteorites, planets, lunar samples, the solar wind, and solar flares since 1960, and advances in understanding systematic trends in values of Aston's packing fraction for the 2,850 nuclides currently known[8] suggest the following modifications to the earlier conclusion of Fowler et al.[7] :

- Local element synthesis may have produced, not just the rare isotopes suggested by Fowler et al.[7], but also bulk material of the solar system and imprinted it with a record of linked chemical and isotopic variations across planetary distances[9].
- Lighter mass elements and the lighter mass isotopes of each element may be enriched at the solar surface. When the photosphere is corrected for this empirical fractionation, the seven most abundant elements in the interior of the Sun seem to be the same ones that comprise 99% of the material in ordinary meteorites. The probability is $< 2 \times 10^{-33}$ that this agreement is fortuitous[9].
- Combined Pu/Xe and U,Th/Pb age dating shows our actinide elements were made in a supernova (SN) explosion at the birth of the solar system, about 5 Ga ago[10].
- Neutron emission from the collapsed SN remnant at the solar core produces $^1$H at the Sun's surface, $^1$H in the solar wind, and >57% of the Sun's energy[11]. Solar fusion of this neutron decay product generates <38% of its energy[11].

Table 1 compares the standard solar model[12] (ssm) with the new solar model[11].

Table 1. Two Models of the Sun

| Properties | Std. Solar Model[12] | New Solar Model[11] |
|---|---|---|
| Origin | The sun formed instantly as a homogeneous body from an interstellar cloud with no mass accretion or mass loss. | The sun formed in a timely manner by accretion of fresh SN debris on the collapsed core of a supernova. |
| Main source of luminosity | Hydrogen-fusion in the core | Energy from a SN core |
| Main nuclear reactions | Hydrogen fusion: <br> 4 $^1$H + 2 e- → <br> $^4$He + 2 v + 27 MeV | a) Neutron emission: <br> $<^1$n$>$ → $^1$n + 10 MeV <br> b) Neutron decay: $^1$n → $^1$H + anti-v + 0.8 MeV <br> c) Hydrogen fusion: <br> 4 $^1$H + 2 e- → <br> $^4$He + 2 v + 27 MeV |
| Energy from H-fusion | ≈ 100 % | < 38 % |
| Solar neutrino flux observed/predicted, excluding oscillations | ≈ 33 % | > 87 % |
| Observable by-products of solar luminosity | Neutrinos from decay of fusion products in the core. | a) Anti-neutrinos from decay of neutrons near the core. <br> b) Neutrinos from fusion product decay near the core. <br> c) H$^+$ ions escape from the surface in the solar wind. |
| Major elements | Hydrogen, helium, carbon | Iron, nickel, oxygen, silicon |
| Comparable meteorites | None. Only about 0.1 % of the sun has the composition of carbonaceous chondrites. | Most. The Sun and ordinary meteorites consist mostly of Fe, O, Si, Ni, S, Mg and Ca. |
| Comparable planets | Gaseous planets, far from sun | Rocky planets, close to sun |

If the ssm is correct, there is a clear deficit of solar neutrinos and the neutral current observed in the SNO experiment[13] likely originates in the Sun. If the new solar model is correct, the solar neutrinos detected seem to represent the bulk (> 87%) of those produced in the Sun and the neutral current observed in the SNO experiment[13] is likely of non-solar origin.

Acknowledgements

This paper is dedicated to the memory of Francis William Aston and Paul Kazuo Kuroda, whose careful measurements and insight into nuclear energy form the basis for this new model of the Sun. The Foundation for Chemical Research, Inc. and the University of Missouri-Rolla supported this research.